# Retrieving sinusoids from nonuniformly sampled data using recursive formulation

Ivan Maric

*Abstract*— A heuristic procedure based on novel recursive formulation of sinusoid (RFS) and on regression with predictive least-squares (LS) enables to decompose both uniformly and nonuniformly sampled 1-d signals into a sparse set of sinusoids (SSS). An optimal SSS is found by Levenberg-Marquardt (LM) optimization of RFS parameters of near-optimal sinusoids combined with common criteria for the estimation of the number of sinusoids embedded in noise. The procedure estimates both the cardinality and the parameters of SSS. The proposed algorithm enables to identify the RFS parameters of a sinusoid from a data sequence containing only a fraction of its cycle. In extreme cases when the frequency of a sinusoid approaches zero the algorithm is able to detect a linear trend in data. Also, an irregular sampling pattern enables the algorithm to correctly reconstruct the under-sampled sinusoid. Parsimonious nature of the obtaining models opens the possibilities of using the proposed method in machine learning and in expert and intelligent systems needing analysis and simple representation of 1-d signals. The properties of the proposed algorithm are evaluated on examples of irregularly sampled artificial signals in noise and are compared with high accuracy frequency estimation algorithms based on linear prediction (LP) approach, particularly with respect to Cramer-Rao Bound (CRB).

## I. Introduction

### A. Problem statement

LET $\{w_k\}_{k=1}^K$ denote a time series, where $w_k \in \Re (k=1,...,K)$ is the $k$th observation obtained at the corresponding time point $t_k$, $\{t_k\}_{k=1}^K$. Suppose a time series representing a finite number of sine waves embedded in noise. Suppose also that a time series may have a nonzero mean value and/or a linear trend. The objective of this paper is spectral analysis and modeling of a time series outlined above and represented by:

$$w_k = o + \kappa t_k + \sum_{n=1}^{N}[A_n \cdot \sin(\omega_n \cdot t_k + \varphi_n)] + s_k, k=1,...,K, \quad (1)$$

This work has been supported by the Croatian Ministry of Science and Education through the project "Computational Intelligence Methods in Measurement Systems" No. 098-0982560-2565 within the Centre of Research Excellence for Data Science and Cooperative Systems.
I.Maric is with the Division of Electronics, Rudjer Boskovic Institute, Bijenicka 54, 10000 Zagreb, Croatia (e-mail: ivan.maric@irb.hr).

where $o$ and $\kappa$ denote the corresponding y-intercept at $t=0$ and the slope of a linear trend line, $A_n$, $\omega_n$ and $\varphi_n$ are the corresponding amplitude, radian frequency and phase of the $n$th sine wave and $s_k$ represents the noise.

### B. Related work

A non-uniform sampling is common to many long-time ground-based astronomical observations including spectra and time series (Lomb, 1976; Scargle, 1982). A number of papers dealing with the decomposition of a time series into a SSS are based on the least-squares spectral analysis and have been published very early. Methods based on the least-squares fit of sinusoids to data are introduced, also known as LS periodogram (LSP) analysis, formulated as LS fitting problem:

$$\min_{\substack{A_n \geq 0 \\ \omega_n \in [0,\omega_{max}] \\ \varphi_n \in [-\pi,\pi]}} \sum_{k=1}^{K}\left[w_k - \sum_{n=1}^{N} A_n \cdot \sin(\omega_n \cdot t_k + \varphi_n)\right]^2, \quad (2)$$

where $\omega_{max}$ denotes maximum expected angular frequency. Frequency estimation methods can be divided into the two main classes: nonparametric and parametric. The nonparametric frequency estimation is based on the Fourier transform and its ability to resolve closely spaced sinusoids is limited by the length of sampled data. On the other hand the parametric approach enables to achieve a higher resolution since it assumes the generating model with known functional form, which satisfies the signal (So et al., 2005).

The earliest nonparametric frequency estimation methods are based on LSP analysis. Barning (1962) used least-squares fitting to calculate the amplitudes of sine waves from the corresponding frequencies selected from periodogram. Vaníček (1969) first proposed successive spectral analysis of equally spaced data and later he extended the analysis to nonuniformly sampled data (Vaníček, 1970). Lomb, (1976) analyzed statistical properties of irregularly spaced data based on periodogram analysis. He has shown that, due to the correlation between noise at different frequencies, noise has less effect on a spectrum than it could be expected. Scargle (1982) studied the use of periodogram with irregularly spaced data. He concluded that periodogram analysis and least-squares fitting of sine waves to data are exactly equivalent. Foster (1995) proposed a sequential method for removing false peaks from power spectra that can be viewed as Matching Pursuit (Mallat, & Zhang, 1993), a general procedure for computing adaptive signal representations which decomposes any signal into a linear expansion of waveforms that are selected from a redundant dictionary of functions. Bourguignon, Carfantan, and Idier, (2007) estimated spectral components from irregularly sampled data. Sparse representation of noisy data is searched for in an arbitrarily large dictionary of complex-valued sinusoidal signals, which can be viewed as Basis Pursuit Denoising problem (Chen, Donoho, & Saunders, 2001). The nonparametric method for spectral analysis of nonuniform sequences of real-valued data named real-valued iterative adaptive approach (RIAA) is proposed by Stoica, Li, and He (2009). It can be interpreted as an iteratively weighted LSP. The method can be used for spectral analysis of general data sequences but is most suitable for zero-mean sequences with discrete spectra. Similar problems, dealing with sparse reconstruction, have been investigated recently in scope of compressed sensing, (Tang et al., 2012; Nichols, Oh, & Willett, 2014; Boufounos et al., 2012; Panahi & Viberg, 2014; Teke, Gurbuz, & Arikan, 2013), illustrating only signal reconstruction errors but not demonstrating that the proposed methods achieve a Cramer–Rao bound, above some SNR threshold, for all the real frequencies embedded in the signal.

Well-known parametric frequency estimation methods are maximum likelihood (ML) (Rife, & Boorstyn, 1976; Bresler & Macovski, 1986), and nonlinear least squares (NLS) (Stoica & Nehorai, 1988) and the methods based on linear prediction (LP) property of sinusoids like Yule–Walker equations (Chan, & Langford, 1982), total least squares, (Rahman, & Yu, 1987), iterative filtering (Li, & Kedem, 1994), MUSIC and ESPRIT (Porat, 2008), weighted least squares (So et al., 2005). Under additive white Gaussian noise the ML and NLS methods are equivalent and achieve Cramer–Rao lower bound (CRLB) asymptotically, but they are computationally demanding. The above mentioned methods, based on LP property, provide suboptimum estimation performance but they are computationally efficient. The parametric methods based on linear prediction (Chan, Lavoie, & Plant, 1981; So, et al., 2005; Dash, & Hasan, 2011; Yang, Xi, & Guo, 2007) enable to retrieve the sinusoids from a uniformly sampled sinusoidal signal in noise when the number of sinusoids in the signal is known *a priori*. So et al. (2005) developed two high accuracy frequency estimators for multiple real sinusoids in white noise based on the LP approach. First, they developed a constrained least squares frequency estimator named reformulated Pisarenko harmonic decomposer (RPHD) and then they improved it through the technique of weighted least squares (WLS) with a generalized unit-norm (WLSun) and monic (WLSm) constraint. The method assumes uniformly sampled data and the number of sinusoids to be known *a priori*.

The heuristic procedure elaborated in this paper is also based on the LP property of a sinusoid and is intended for recovery of frequency-sparse signals in noise. It can be used in signal processing, machine learning and expert and intelligent systems to facilitate solving the classification, diagnosis, monitoring or process control tasks needing analysis and parsimonious representation of signals, including the signals in technical systems, bio-signals, astronomical observations,
2

which decomposes any signal into a linear expansion of waveforms that are selected from a redundant dictionary of functions. Bourguignon, Carfantan, and Idier, (2007) estimated spectral components from irregularly sampled data. Sparse representation of noisy data is searched for in an arbitrarily large dictionary of complex-valued sinusoidal signals, which can be viewed as Basis Pursuit Denoising problem (Chen, Donoho, & Saunders, 2001). The nonparametric method for spectral analysis of nonuniform sequences of real-valued data named real-valued iterative adaptive approach (RIAA) is proposed by Stoica, Li, and He (2009). It can be interpreted as an iteratively weighted LSP. The method can be used for spectral analysis of general data sequences but is most suitable for zero-mean sequences with discrete spectra. Similar problems, dealing with sparse reconstruction, have been investigated recently in scope of compressed sensing, (Tang et al., 2012; Nichols, Oh, & Willett, 2014; Boufounos et al., 2012; Panahi & Viberg, 2014; Teke, Gurbuz, & Arikan, 2013), illustrating only signal reconstruction errors but not demonstrating that the proposed methods achieve a Cramer–Rao bound, above some SNR threshold, for all the real frequencies embedded in the signal.

Well-known parametric frequency estimation methods are maximum likelihood (ML) (Rife, & Boorstyn, 1976; Bresler & Macovski, 1986), and nonlinear least squares (NLS) (Stoica & Nehorai, 1988) and the methods based on linear prediction (LP) property of sinusoids like Yule–Walker equations (Chan, & Langford, 1982), total least squares, (Rahman, & Yu, 1987), iterative filtering (Li, & Kedem, 1994), MUSIC and ESPRIT (Porat, 2008), weighted least squares (So et al., 2005). Under additive white Gaussian noise the ML and NLS methods are equivalent and achieve Cramer–Rao lower bound (CRLB) asymptotically, but they are computationally demanding. The above mentioned methods, based on LP property, provide suboptimum estimation performance but they are computationally efficient. The parametric methods based on linear prediction (Chan, Lavoie, & Plant, 1981; So, et al., 2005; Dash, & Hasan, 2011; Yang, Xi, & Guo, 2007) enable to retrieve the sinusoids from a uniformly sampled sinusoidal signal in noise when the number of sinusoids in the signal is known *a priori*. So et al. (2005) developed two high accuracy frequency estimators for multiple real sinusoids in white noise based on the LP approach. First, they developed a constrained least squares frequency estimator named reformulated Pisarenko harmonic decomposer (RPHD) and then they improved it through the technique of weighted least squares (WLS) with a generalized unit-norm (WLSun) and monic (WLSm) constraint. The method assumes uniformly sampled data and the number of sinusoids to be known *a priori*.

The heuristic procedure elaborated in this paper is also based on the LP property of a sinusoid and is intended for recovery of frequency-sparse signals in noise. It can be used in signal processing, machine learning and expert and intelligent systems to facilitate solving the classification, diagnosis, monitoring or process control tasks needing analysis and parsimonious representation of signals, including the signals in technical systems, bio-signals, astronomical observations, etc. The proposed algorithm enables to retrieve the sinusoids from either uniformly or nonuniformly sampled data. In order to adapt it to nonuniform sampling we first reformulate the LP property of a sinusoid and we named it a recursive formulation of a sinusoid (RFS). Then we formulate a sinusoidal model based on RFS and the corresponding procedures for the estimation of RFS parameters based on the minimization of LS error. By combining the RFS approach with the well-known methods for the estimation of the number of sinusoids in noise the proposed procedure enables to retrieve the sinusoids iteratively, one at a time, and to determine the order of the generating model. The proposed method assumes neither a zero mean sequence nor the number of sinusoids in a signal to be known *a priori*. The accuracy of the frequency estimation procedure proposed in this paper is compared with very high accuracy of frequency estimation obtained by LP approach reported by So et al. (2005). For a frequency-sparse signal the computational complexity of both methods is comparable, $O(K^3)$.

C. *Methods for detection of the number of sinusoids*

Most parametric methods for detection of sinusoids corrupted with noise minimize the sum of a data fit (likelihood) term and the complexity penalty term where the penalty term is usually derived via Akaike information criterion (AIC) (Akaike, 1974), Bayesian information criteria (BIC) (Schwarz, 1978) or minimum description length (MDL) (Rissanen, 1978). A review of information criterion rules for model-order selection with the summary of necessary steps used to adapt a rule to a specific problem is given in Stoica and Selen (2004). In this paper the attention is restricted to efficient detection criteria (EDC) type estimators (Djurić, 1996, 1998; Nadler & Kontorovich, 2011). EDC type estimators determine the number of sinusoids by minimizing:

$$\hat{M} = \arg \min_{M=0,1,2...} -\ln L(\hat{\beta}_M, \mathbf{w}) + MC_K, \quad (3)$$

where $\mathbf{w}$ is the observed time series of length $K$, $\hat{\beta}_M$ are parameter estimates of a model of order $M$, $L(\hat{\beta}_M, \mathbf{w})$ is the corresponding likelihood term and $C_K$ is the model-complexity penalty term that captures the dependency of the penalty on the number of samples $K$. For the unknown noise level the log-likelihood term in (3) can be approximated by:

$$\ln L(\hat{\beta}_M, \mathbf{w}) = -\frac{K}{2} \ln \left\{ \sum_{k=1}^{K} \left[ w_k - P_{M,k}(\hat{\beta}_M) \right]^2 \right\}^{0.5}, \quad (4)$$

where $P_{M,k}(\hat{\beta}_M)$ denotes the approximation of $w_k$ at time point $t_k$ made by a model of order $M$. By substituting (4) for log-likelihood in (3) we obtain:

$$\hat{M} = \arg \min_{M=0,1,2,...} \frac{K}{2} \ln \left\{ \sum_{k=1}^{K} \left[ w_k - P_{M,k}(\hat{\beta}_M) \right]^2 \right\}^{0.5} + MC_K. \quad (5)$$

By considering a Bayesian formulation and selecting the model with maximum *a posteriori* probability (MAP) criterion for sinusoids with unknown frequencies amplitudes and phases Djurić (1996) derived the following penalty term

$$C_K = (5M/2)\ln K \quad (6)$$

and he concluded that the parameters that can be determined more precisely should receive stronger penalty. Nadler and Kontorovich (2011) proposed the estimator inspired by ideas from extreme value theory (EVT) and the maxima of stochastic fields with the following penalty term

$$C_K = \ln K + \frac{1}{2}\ln\ln K - \frac{1}{2}\ln\left(\frac{3\alpha^2}{\pi}\right), \quad (7)$$

where $\alpha \ll 1$ denotes a confidence level chosen by the user, typically $\alpha \leq 0.005$. They recommend the generalized likelihood ratio test (GLRT)

$$\ln\left(\frac{L(\hat{\beta}_M, \mathbf{w})}{L(\hat{\beta}_{M-1}, \mathbf{w})}\right) > C_K \quad (8)$$

to determine the number of sinusoids ($M$).

Next section elaborates the RFS and the RFS-based regression procedures and the corresponding RFS-based algorithm (RFSA) used to retrieve the sinusoids from nonuniformly sampled data. In Section III the frequency estimation accuracy of the proposed procedure is compared with high accuracy LP approach (So et al., 2005). Also the results of spectral analysis of a couple of nonuniformly sampled 1-d signals are given to illustrate the properties of the proposed method.

## II. TIME SERIES ANALYSIS BY RFS

This section presents a novel RFS based procedure for retrieving the sinusoids from unevenly spaced data. The proposed procedure is able to precisely estimate the total number and the parameters of SSS from uniformly and nonuniformly sampled sinusoidal signal in noise. It can discover a cyclical pattern with linear trend in data (e.g. excitation signals in AC voltammetry, atmospheric Carbon Dioxide data) or to retrieve an undersampled sinusoid or a low frequency sinusoid in cases when only a fraction of its cycle is covered by a time series. The procedure is based on minimization of accumulated prediction error using $\ell^2$-norm. The frequencies from a predefined set of frequencies are optimized individually by LM (Levenberg, 1944; Marquardt, 1963) in order to obtain the parameters of a sinusoid which best minimizes the predictive LS error. Next, LM optimization is used to fine tune the RFS parameters of all most dominant sine waves found until then, resulting with a decomposition of a time series into an optimal set of sinusoidal components. In order to determine the cardinality of a SSS, the procedure combines the criteria for the detection of the number of sinusoids embedded in noise (Nadler & Kontorovich, 2011; Djurić, 1996; Djurić, 1998) (see Section I–C).

The idea of RFS in nonuniform and uniform sampling case and its adaptation for straight line approximation is given below, followed by a reformulation of the LS fitting problem (2) in terms of sine wave representation by a RFS. Next, the procedure for calculating pairs of initial samples of the sinusoids is presented, then the elaboration of the LM optimization of RFS parameters is given and finally the explanation of RFS model order estimation procedure, which rounds up the methodology. The section concludes with the description of the RFSA algorithm.

### A. Recursive Formulation of a Sinusoid and a Straight Line
#### 1) Nonuniform sampling case

A sinusoid $y_m = A_m \sin(\omega_m t + \varphi_m)$ can be predicted by using a simple RFS (see Appendix A), which relates any sample of a sinusoid with its two referent samples, e.g. two initial samples:

$$y_{m,k} = a_{m,k} y_{m,2} + b_{m,k} y_{m,1} \quad (9)$$

where $m$ denotes a sine wave with the corresponding radial frequency ($\omega_m$), amplitude ($A_m$) and phase ($\varphi_m$), $y_{m,k}$ denotes the predicted magnitude of the sine wave at time point $t_k$, $y_{m,1}$ and $y_{m,2}$ represent two initial samples obtained at the corresponding time points $t_1$ and $t_2$, $a_{m,k}$ and $b_{m,k}$ are time and frequency dependent coefficients defined as

$$a_{m,k} = \frac{\sin(\omega_m \tau_{k,1})}{\sin(\omega_m \tau_{2,1})} \quad (10)$$

and

$$b_{m,k} = -\frac{\sin(\omega_m \tau_{k,2})}{\sin(\omega_m \tau_{2,1})} \quad (11)$$

with $\tau_{j,i} = t_j - t_i$ representing the difference in seconds between the time points of $j$th and $i$th sample from the sequence of samples and $\omega_m$ denoting angular frequency of the sine wave in rad/s. Note that the radian frequency and the two initial samples ($\omega_m, y_{m,1}, y_{m,2}$) are the parameters of RFS (9), which completely specify the corresponding sinusoid.

If $\omega_m \to 0$, (10) and (11) can be replaced by $a_{m,k} \approx \tau_{k,1}/\tau_{2,1}$ and $b_{m,k} \approx -\tau_{k,2}/\tau_{2,1}$, respectively. Substituting $\tau_{k,2} = \tau_{k,1} - \tau_{2,1}$ in $b_{m,k}$ and then $a_{m,k}$ and $b_{m,k}$ in (10) the following approximate equation is obtained:

$$y_{m,k} \underset{\omega_m \to 0}{\approx} \frac{y_{m,2} - y_{m,1}}{\tau_{2,1}} \tau_{k,1} + y_{m,1}, \quad (12)$$

which can be recognized as a recursive formulation of an arbitrary straight line. Hence, for the given angular frequency

$\omega_m \in [0, \omega_{max}]$ and the two initial samples $y_{m,1}$ and $y_{m,2}$ with the corresponding time points $t_1$ and $t_2$, any sample $y_{m,k}$ of a sine wave, including a straight line as a special case when $\omega=0$, can be accurately predicted at the time point $t_k$ by using (9)–(12).

*2) Uniform sampling case*

In case of uniform sampling the coefficients (10) and (11) become $a_{m,k} = \frac{\sin[(k-1)\omega_m T]}{\sin(\omega_m T)}$ and $b_{m,k} = -\frac{\sin[(k-2)\omega T]}{\sin(\omega T)}$, where $T$ denotes a sampling period and the coefficients are now calculated recursively by adapting Chebyshev multiple angle formula, i.e.

$$a_{m,k} = x_m a_{m,k-1} + b_{m,k-1} \quad (13)$$

$$b_{m,k} = -a_{m,k-1} \quad (14)$$

where $a_{m,1}=0$, $b_{m,1}=1$, and

$$x_m = 2\cos(\omega_m T). \quad (15)$$

Note that in uniform sampling case the parameter $x_m \in [-2,2]$ is equivalent to frequency parameter $\omega_m$ and the calculation of the coefficients (13) and (14) is reduced to FP multiplications and additions only. After the parameter $x_m$ is estimated, it can be easily converted into the frequency

$$\omega_m = \cos^{-1}(x_m/2)/T. \quad (16)$$

Note that Eqs. (13) – (16) are valid for both, a sinusoid and an arbitrary straight line ($\omega_m =0 \Leftrightarrow x_m=2$).

*B. Reformulation of LS Fitting Problem*

After substituting the RFS (9) for each sinusoidal component, including a trend line (12), the time series (1) can be represented by the following relation:

$$w_k = \sum_{m=1}^{M}(a_{m,k} y_{m,2} + b_{m,k} y_{m,1}) + s_k, k = 1,...,K, \quad (17)$$

where $a_{m,k}$ and $b_{m,k}$ denote the coefficients (10) and (11) of the $m$th RFS (9) and $N$ in (1) is replaced by $M=N+1$ in (17) because additional RFS in (17) is used to represent a linear trend in data (1). Recall that recursive formulation of a straight line (12) is a special case of RFS (9) when the frequency of a sinusoid approaches zero. Note that two initial samples $y_{m,1}$ and $y_{m,2}$ in each RFS $m=1,...,M$ in (17) and the corresponding angular frequency $\omega_m$, which affects the coefficients $a_{m,k}$ (10) and $b_{m,k}$ (11) are all considered independent variables. Hence, the sinusoidal signal can be restored from noisy data sequence (17) if the initial samples and the frequencies of the corresponding sinusoids can be estimated. The LS fitting problem (2) is re-formulated in the following way:

$$E_M = \min_{\substack{y_{m,1}\in\langle-\infty,+\infty\rangle \\ y_{m,2}\in\langle-\infty,+\infty\rangle \\ \omega_m \in [\omega_{min},\omega_{max}] \text{ or } x_m \in [-2,2]}} \sum_{k=1}^{K}\left[w_k - \sum_{m=1}^{M}(a_{m,k} y_{m,2} + b_{m,k} y_{m,1})\right]^2, \quad (18)$$

where $\omega_{min}$ and $\omega_{max}$ denote the corresponding lower and upper bound for possible angular frequencies $\omega_m \in [\omega_{min}, \omega_{max}]$ and $M$ is the number of detected sinusoids. The LS fitting of RFS to data (18) is different from (2) since it employs prediction rather than approximation to estimate RFS parameters $\boldsymbol{\beta}$, $\{\boldsymbol{\beta}_m\}_{m=1}^{M} = \{\omega_m, y_{m,1}, y_{m,2}\}_{m=1}^{M}$, or $\{\boldsymbol{\beta}_m\}_{m=1}^{M} = \{x_m, y_{m,1}, y_{m,2}\}_{m=1}^{M}$ in the uniform sampling case, by minimizing the error function based on predictive least squares (Rissanen, 1986). If it would be possible to estimate the RFS parameters, by solving the LS fitting problem (18), then it should also be possible to reconstruct the time series (17) or (1) as well as to calculate the amplitudes and phases of all sine waves (see Appendix C).

To solve the LS fitting problem (18) the following procedures are necessary:
1. Calculation of initial samples of sine waves that best minimize (18) for the given angular frequencies (see Section II–D).
2. Optimization of parameters (frequencies and initial samples) of multiple RFS by LM algorithm (see Section II–E).

This new formulation, when applied to nonuniformly sampled data representing multiple superimposed oscillations (MSO), enables to recover the sinusoid even if sampled data represent only a fraction of its cycle as well as to recover the under-sampled sinusoid whose frequency might be higher than the Nyquist frequency defined by the Nyquist–Shannon sampling theorem (Shannon, 1998). In case of nonuniform sampling the Nyquist frequency can be pushed very high (Eyer & Bartholdy, 1999; Koen, 2006). For the given angular frequency $\omega_x$, the proposed procedure maps all nonuniformly spaced angles into the same normalized sine wave period $[0, 2\pi]$, using the relation $\mathrm{mod}_{2\pi}(\omega_x t_k + \varphi_x)$, thus artificially shortening the average sampling period. The design of optimal sampling pattern is beyond the scope of this paper.

*C. RFS Model Estimation*

Let $\{w_k\}_{k=1}^{K}$ be a nonuniform time series with the corresponding time points $\{t_k\}_{k=1}^{K}$ represented by (1) or equivalently by (17). A solution to LS fitting problem (18) is the following RFS model of a time series:

$$P_{M,k}(\boldsymbol{\beta}) = \sum_{m=1}^{M} y_{m,k} = \sum_{m=1}^{M}(a_{m,k} y_{m,2} + b_{m,k} y_{m,1}), k=1,...,K \quad (19)$$

where $P_{M,k}(\boldsymbol{\beta})$ denotes the predicted value of $k$th sample of a time series represented by a superposition of $M$ RFS. The time and frequency dependent coefficients $a_{m,k}$ and $b_{m,k}$ are defined by (10) and (11), respectively. To derive the model (19) we need to estimate the RFS parameters $\boldsymbol{\beta}$. The corresponding algorithm (see Section II–G) iteratively estimates the most dominant sinusoids in the signal. It uses predefined frequencies to find the suboptimal RFS parameters close to real RFS parameters (Section II–D), then it optimizes RFS parameters (Section II–E) trying to solve the LS fitting problem (18). The procedure starts with estimation and optimization of RFS parameters of the first most dominant sinusoid, than continues with the estimation and optimization of the RFS parameters of the two most dominant sinusoids etc. The procedure combines EDC estimators to select the model order.

*D. Calculation of Initial Samples of Sine Waves*

This section details the calculation procedure, which enables direct solution to (18) for $M$ predefined frequencies. Given the frequencies, $\{\omega_m\}_{m=1}^M$, or $\{x_m\}_{m=1}^M$ in the uniform sampling case, the LS prediction error (18) has to be minimized with respect to initial samples of sinusoids $\boldsymbol{\alpha}$, $\{\alpha_m\}_{m=1}^M = \{y_{m,1}, y_{m,2}\}_{m=1}^M$, where the coefficients $a_{m,k}$ and $b_{m,k}$ in (18) are calculated by (10) and (11) using the preselected frequencies $\{\omega_m\}_{m=1}^M$, or in case of uniform sampling by (13) and (14) using the preselected parameters (15), $\{x_m\}_{m=1}^M$. Note that $\boldsymbol{\alpha}$ is a subset of $\boldsymbol{\beta}$, $\boldsymbol{\alpha} \subseteq \boldsymbol{\beta}$. After setting the partial derivatives of (18) with respect to initial samples equal to zero a set of $2M$ simultaneous linear equations in matrix form is obtained:

$$(\mathbf{J}^T\mathbf{J})\boldsymbol{\alpha} = \mathbf{J}^T\mathbf{w}, \qquad (20)$$

where $\mathbf{w}$ is a time series vector, $\boldsymbol{\alpha}$ is a parameter vector to calculate and $\mathbf{J}$ is a Jacobian matrix of time series prediction model $\mathbf{P}_{M,k}(\boldsymbol{\beta})$ (16) with respect to $\boldsymbol{\alpha}$, $\boldsymbol{\alpha} \subseteq \boldsymbol{\beta}$, (see Appendix B). Eq. (20) can be solved directly for $\boldsymbol{\alpha}$.

*E. Optimization of RFS parameters by LM algorithm*

The parameter vector $\boldsymbol{\beta}$ of the RFS model (19), obtained in Section II–C, can be optimized by LM algorithm (Levenberg, 1944; Marquardt, 1963) in order to further minimize the LS error (18). LS fitting problem (18), adapted for LM optimization takes the form of a nonlinear error function:

$$E_{LM}(\boldsymbol{\beta}+\boldsymbol{\delta}) = \sum_{k=1}^{K}[w_k - P_{M,k}(\boldsymbol{\beta}+\boldsymbol{\delta})]^2 \approx$$
$$\approx \sum_{k=1}^{K}\left[w_k - \sum_{m=1}^{M}\left(y_{m,k} + \frac{\partial y_{m,k}}{\partial \omega_m}\delta\omega_m + \frac{\partial y_{m,k}}{\partial y_{m,1}}\delta y_{m,1} + \frac{\partial y_{m,k}}{\partial y_{m,2}}\delta y_{m,2}\right)\right]^2, \quad (21)$$

where $\boldsymbol{\delta}$ denotes the parameter increment vector to calculate, $\{\delta_m\}_{m=1}^M = \{\delta\omega_m, \delta y_{m,1}, \delta y_{m,2}\}_{m=1}^M$ or $\{\delta_m\}_{m=1}^M = \{\delta x_m, \delta y_{m,1}, \delta y_{m,2}\}_{m=1}^M$, and $\boldsymbol{\beta}$ is the parameter vector to optimize. Note that in case of uniform sampling the parameter $x_m$ has to be substituted for $\omega_m$ in (21) and in $\boldsymbol{\beta}$ and $\boldsymbol{\delta}$. After setting the partial derivatives of (21) with respect to increments equal to zero and after introducing an adjustable nonnegative damping factor $\gamma$ and the diagonal matrix of $\mathbf{J}^T\mathbf{J}$, where $\mathbf{J}$ is a Jacobian matrix of a time series model $\mathbf{P}_{M,k}(\boldsymbol{\beta})$ (19) with respect to parameters $\boldsymbol{\beta}$, a well-known LM equation in matrix notation is obtained:

$$\left[\mathbf{J}^T\mathbf{J} + \gamma \cdot diag\left(\mathbf{J}^T\mathbf{J}\right)\right]\boldsymbol{\delta} = \mathbf{J}^T[\mathbf{w}-\mathbf{P}(\boldsymbol{\beta})], \qquad (22)$$

summarizing a set of $3M$ linear equations with $3M$ unknowns ($\boldsymbol{\delta}$). A more detailed description of (22) is given in Appendix D. Over a preset number of steps $L$ the LM algorithm successively modifies the parameter vector ($\boldsymbol{\beta}_{l+1}=\boldsymbol{\beta}_l+\boldsymbol{\delta}_l$) by the $l$th instance of the increment vector ($\boldsymbol{\delta}$), obtained from (22). In nonuniform sampling case the partial derivatives of $y_{m,k}$ with respect to $\omega_m$, $y_{m,1}$ and $y_{m,2}$ in (21) are derived from (9) after substituting (10) and (11) for $a_{m,k}$ and $b_{m,k}$, respectively:

$$\frac{\partial y_{m,k}}{\partial y_{m,1}} = -\frac{\sin(\omega_m\tau_{k,2})}{\sin(\omega_m\tau_{2,1})}, \qquad (23)$$

$$\frac{\partial y_{m,k}}{\partial y_{m,2}} = \frac{\sin(\omega_m\tau_{k,1})}{\sin(\omega_m\tau_{2,1})}. \qquad (24)$$

$$\frac{\partial y_{m,k}}{\partial \omega_m} = \frac{\tau_{k,1}\cos(\omega_m\tau_{k,1})\sin(\omega_m\tau_{2,1})}{[\sin(\omega_m\tau_{2,1})]^2}y_{m,2}$$
$$- \frac{\tau_{2,1}\sin(\omega_m\tau_{k,1})\cos(\omega_m\tau_{2,1})}{[\sin(\omega_m\tau_{2,1})]^2}y_{m,2}$$
$$- \frac{\tau_{k,2}\cos(\omega_m\tau_{k,2})\sin(\omega_m\tau_{2,1})}{[\sin(\omega_m\tau_{2,1})]^2}y_{m,1} \qquad (25)$$
$$+ \frac{\tau_{2,1}\sin(\omega_m\tau_{k,2})\cos(\omega_m\tau_{2,1})}{[\sin(\omega_m\tau_{2,1})]^2}y_{m,1}$$

The partial derivatives (23) – (25) are derived by assuming the mutual independence of the RFS parameters:

$$\frac{\partial y_{m,1}}{\partial \omega_m} = \frac{\partial y_{m,2}}{\partial \omega_m} = \frac{\partial y_{m,1}}{\partial y_{m,2}} = \frac{\partial y_{m,2}}{\partial y_{m,1}} = 0 \text{ and } \frac{\partial y_{m,1}}{\partial y_{m,1}} = \frac{\partial y_{m,2}}{\partial y_{m,2}} = 1. \quad (26)$$

In uniform sampling case the partial derivatives of $y_{m,k}$ with respect to $x_m$, $y_{m,1}$ and $y_{m,2}$ in (21) are derived from (9) after substituting (13) and (14) for $a_{m,k}$ and $b_{m,k}$, respectively:

$$\frac{\partial y_{m,k}}{\partial y_{m,1}} = -a_{m,k-1}, \qquad (27)$$

$$\frac{\partial y_{m,k}}{\partial y_{m,2}} = x_m a_{m,k-1} + b_{m,k-1}. \qquad (28)$$

$$\frac{\partial y_{m,k}}{\partial x_m} = \frac{\partial a_{m,k}}{\partial x_m}y_{m,2} + \frac{\partial b_{m,k}}{\partial x_m}y_{m,1}, \qquad (29)$$

where

$$\frac{\partial a_{m,k}}{\partial x_m} = \left( a_{m,k-1} + x_m \frac{\partial a_{m,k-1}}{\partial x_m} \right) + \frac{\partial b_{m,k-1}}{\partial x_m}, \quad (30)$$

$$\frac{\partial b_{m,k}}{\partial x_m} = -\frac{\partial a_{m,k-1}}{\partial x_m}, \quad (31)$$

and where $\frac{\partial a_{m,1}}{\partial x_m} = \frac{\partial b_{m,1}}{\partial x_m} = 0$.

Note that the components of partial derivatives $a_{m,k}$, $b_{m,k}$, $\partial a_{m,k}/\partial x_m$ and $\partial b_{m,k}/\partial x_m$ can be calculated completely recursively by using FP multiplications and additions only.

*F. Error minimization and RFS model order selection*

The RFS approach can be efficiently combined with well-known criteria for detection of the number of sinusoids embedded in noise. The number of sinusoids is estimated by following the procedure outlined in Stoica, Li, and He (2009). Let

$$\breve{\boldsymbol{\beta}}_M = \{\breve{\omega}_m, \breve{y}_{m,1}, \breve{y}_{m,2}\}_{m=1}^M \quad (32)$$

denote the RFS parameters of the corresponding $M$ sinusoids used to approximate the time series by the RFS model (19). The corresponding errors (18), due to the approximation of a time series by a certain number $M=1,2,\ldots,M_{max}$ of superimposed sinusoids, are arranged in a decreasing order of their values:

$$E_1 \geq E_2 \geq \ldots \geq E_{M_{max}}. \quad (33)$$

Note that the RFS parameters (32) obtained in the preceding steps are optimized in each succeeding step by LM ((21) and (22)) with the aim to minimize the error (18).

Under the idealizing assumptions that a time series consists of a finite number of sinusoidal components and of normal white noise, and that (18) represents maximum likelihood (ML) estimates of frequencies and initial samples of $M_{max}$ such sinusoidal components, the $EDC(M)$ is used to select $M$ in (5), where $EDC(M)$ is obtained after substituting (19) for $P_{M,k}(\hat{\beta}_M)$ in (5). If a time series consists of white noise only, $\{w_k\}=\{s_k\}$, then according to (5) $M=0$ is selected with the corresponding $EDC(0)$:

$$EDC(0) = \frac{K}{2} \ln \left\{ \sum_{k=1}^{K} [w_k]^2 \right\}^{0.5} \quad (34)$$

Note that in case of uniform sampling the parameter $x_m$ needs to be optimized, instead of $\omega_m$, to minimize the error.

*G. RFSA*

This section describes the algorithm (RFSA) for decomposition of a time series into an optimal SSS. The basic steps of the algorithm are outlined in Table I. The procedure is iterative with the corresponding initial guess: $\hat{M}=0$, $\breve{\boldsymbol{\beta}}_0 = \{\ \}$ and $E_0=EDC(0)$ defined by (34). In each succeeding cycle ($M=1,2,\ldots,M_{max}$) a set of predefined trial frequencies $\{\omega_j\}$, $j=1,\ldots,J$ is used to estimate the current ($M$th) most dominant frequency in a time series. Each trial frequency is separately appended to the set of most dominant frequencies obtained in the preceding cycle and each time (20) is solved to estimate the initial samples of $M$ sine waves that best minimize (18). The best obtained set of RFS parameters (including the frequencies) is then optimized by LM (22) in order to further decrease the prediction error (21). In each cycle the RFS parameters $\breve{\boldsymbol{\beta}}_M$ that best minimize (21) are saved along with the corresponding $EDC(M)$ used in (5). When the condition $M \geq M_{max}$ has been satisfied, the model order $\hat{M}$ (5) and the corresponding RFS parameters $\breve{\boldsymbol{\beta}}_{\hat{M}}$ are determined. The GLRT stopping criterion (8) slightly increases the probability of underestimation of the number of sinusoids in high noise and is therefore not embedded in RFSA. Note that the product of angular frequency and sampling interval $\omega_m\tau_{2,1}$ may cause an overflow error when calculating (10), (11), (23), (24) and (25). To prevent the possible errors the following constraint is implemented in software: if $\sin|\omega_m\tau_{2,1}| < \delta$, where $|\delta|=10^{-12}$, then set $\sin(\omega_m\tau_{2,1})$ equal to $-\delta$ or $+\delta$ depending on the negative or positive sign of $\sin(\omega_m\tau_{2,1})$, respectively. The RFSA algorithm outlined in Table 1 is the same for uniform sampling case except that parameter $x$ (15) has to be optimized instead of frequency $\omega$ using the corresponding parameter range limits $x_{min}=-2$ and $x_{max}=2$ instead of $\omega_{min}$ and $\omega_{max}$, respectively.

TABLE I
DECOMPOSITION OF TIME SERIES INTO THE OPTIMAL SSS BY RFSA

**Input**
 time series $\{w_k\}$, lowest ($\omega_{min}$) and highest ($\omega_{max}$) expected radian frequency, total number of trial frequencies ($J$), maximum number of sinusoids ($M_{max}$), maximum number of LM optimization steps ($L$)
**Initialization**
 $\hat{M}=0$, $\breve{\boldsymbol{\beta}}_0=\{\ \}$, $e_0 = \sum w_k^2$
**Iteration – main loop**
 **For** $M=1$ to $M_{max}$ step 1
  $e_M=e_{M-1}$
  **For** $j=1$ to $J$ step 1
   $\omega_j=\omega_{min}+(j-1)\cdot(\omega_{max}-\omega_{min})/(J-1)$
   $\boldsymbol{\beta}_M = \breve{\boldsymbol{\beta}}_{M-1} + \{\omega_j,0,0\}$
   Solve $E_M(\boldsymbol{\beta}_M)$ for $\boldsymbol{\alpha}_M \subseteq \boldsymbol{\beta}_M$ (20)
   If $e_M > E_M(\boldsymbol{\beta}_M)$ Then $e_M = E_M(\boldsymbol{\beta}_M)$ and $\breve{\boldsymbol{\beta}}_M = \boldsymbol{\beta}_M$
  **Next** $j$
  Optimize $\breve{\boldsymbol{\beta}}_M$ by LM (22) in up to $L$ steps to minimize $E_{LM}$, (21)
  Calculate and save the corresponding $EDC(M)$, (5).
 **Next** $M$
**Results**
 Return the model order $\hat{M}$ (5) and the RFS parameters $\breve{\boldsymbol{\beta}}_{\hat{M}}$ (32)

To retrieve each new sinusoid the algorithm has to solve (20) and (22) repeatedly. For M<<K, the complexity of (20) and (22), when estimating $M$ sinusoids, is $O(MK^2)$. Since (20) has to be solved $J$ times and (22) $L$ times to retrieve each new most dominant sinusoid, the overall complexity of the algorithm is $O((J+L)M_{max}K^2)$, where $J$, $L$ and $M_{max}$ are the preset numbers of trial (grid) frequencies, LM optimization steps and maximum expected sinusoidal components, respectively. The complexity can be reduced significantly if (20) is solved for all $J$ trial frequencies in parallel i.e. $O(L \cdot M_{max}K^2)$.

## III. RESULTS OF TIME SERIES ANALYSIS BY RFSA

This section describes the results of analysis and modeling of an irregularly sampled MSO. Examples of artificial MSO embedded in noise are given. The accuracy of frequency estimation of RFSA will be compared with high accuracy frequency estimation algorithms based on LP approach (So et al., 2005) in Section III–A, and in the succeeding sections it will be demonstrated how RFSA can efficiently recover under-sampled sinusoid and a sinusoid represented by a fraction of its cycle (Section III–B), two closely spaced sinusoids with linear trend (Section III–C), three closely spaced sinusoids (Section III–D), and 10 sinusoids (Section III–E). Nonuniformly sampled signals with additive noise are considered in all examples. The results are obtained by minimizing EDC (5) with respect to $M$ and by applying the two model-complexity penalizations: MAP (6) and EVT (7). Some general remarks are given in Section III–F.

The CRB for irregular sampling is hard to calculate. It was shown experimentally (Larsson & Larsson, 2002) that CRB is practically the same, but not identical, for different sampling schemes having the same average sampling interval. Hence, the CRB for uniform sampling can be used to approximate the CRB for nonuniform sampling if the average sampling interval of the nonuniform sampling pattern is equal to sampling interval in uniform sampling. To approximate the bound on the frequency for the case of nonuniform sampling the CRB (Porat, 2008, page 265), can be rewritten in the following way:

$$CRB(\omega_m) \approx \frac{24\sigma^2}{K^3 A_m^2 \bar{\tau}^2}\left[1 + \frac{3\cos(2\varphi_m)\sin(K\omega_m)}{K\sin(\omega_m)}\right], \quad (35)$$

where $\bar{\tau}$ denotes the mean sampling interval in seconds and $\sigma^2$ is the noise variance.

The signal-to-noise ratio (SNR) for $m$th sinusoid is defined as $A_m^2/2\sigma^2$ or in dB units:

$$SNR_m = 10\log_{10}\left(\frac{A_m^2}{2\sigma^2}\right) \text{ dB}, \quad (36)$$

where $A_m$ is the amplitude of the $m$th sinusoid. The additive noise is white with zero mean. In all figures the SNR is given with respect to a sinusoid with amplitude $A=1$, i.e. $SNR=1/(2\sigma^2)$, except in figures in Section III–A, where $A=2^{0.5}$ and $SNR=1/\sigma^2$. In all examples the maximum number of LM optimization steps is 30 and the LM damping factor is 1.5. The RFSA is coded in Visual C and executed on Intel®Xeon® CPU E5420 @ 2.50 GHz. To illustrate the computational complexity of the procedure the maximum computation time needed to decompose a single time series is given in each example.

### A. Comparing RFSA with high accuracy frequency estimation algorithm

The accuracy of frequency estimation of the RFSA is compared with LP-based high accuracy frequency estimation algorithm proposed by So et al., (2005). The algorithms have approximately equal computational complexity. The results obtained by RFSA are compared with the results published by

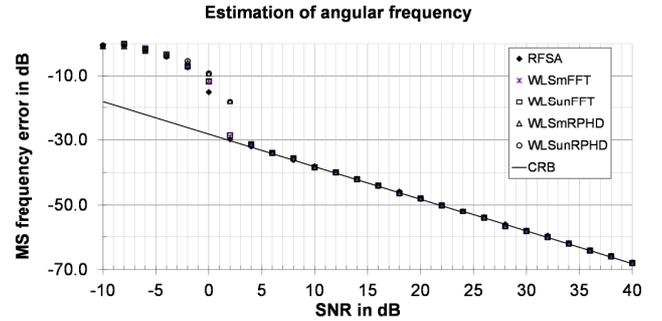
Fig. 1. MSE of the frequency versus SNR obtained by RFSA from irregular data sequences ($K$=20) using EVT model order estimator (7) with $\alpha$=0.1% and 20 trial frequencies.

So et al. (2005). Fig 1 shows a mean squared error (MSE) of frequency of a single sinusoid $y=2^{0.5}sin(0.3\pi)$ in white Gaussian noise obtained from uniformly sampled data ($T$=1s) with $K$=20 samples. SNR values in the range [–10, 40] dB are considered in this experiment. For each SNR value 1000 Monte Carlo (MC) trials are performed. The frequency interval $f\epsilon[0,0.5]$ Hz ($f\epsilon[0,\pi]$ rad/s) is used in RFSA with 0.5/19 Hz ($\pi$/19 rad/s) as a step of a frequency grid ($J$=20 frequencies) and the maximum preset order of a model is

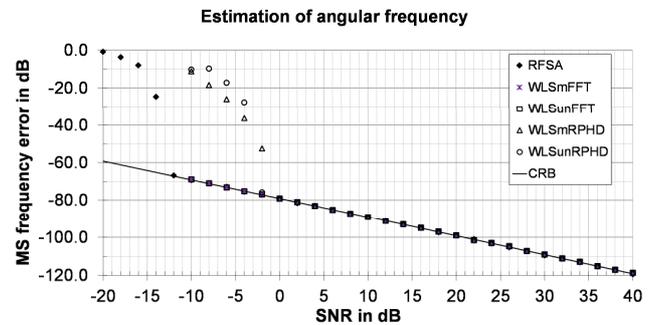
Fig. 2. MSE of the frequency versus SNR obtained by RFSA from irregular data sequences ($K$=1000) using EVT model order estimator (7) with $\alpha$=0.5% and 20 trial frequencies.

$M_{max}=4$. EVT penalization (7) is used with confidence level $\alpha =0.1\%$. From Fig 1 it can be seen that MSE obtained by RFSA is almost identical to MSE obtained by WLS with monic and unit norm constraints when initiated by FFT (WLSmFFT and WLSunFFT) or RPHD (WLSmRPHD and WLSunRPHD). All methods have SNR threshold at about 2dB and attained the CRB for sufficiently high SNR conditions. Note that, unlike the methods published by So et al. (2005), RFSA does not need the number of sinusoids in the signal to be known a priori and it selected correct model order ($M=1$) in the entire range of SNR. Maximum execution time of RFSA for a single trial was 0.385 s. The same test was repeated for K=1000. The frequency interval $f\epsilon[0,0.5]$ Hz ($f\epsilon[0,\pi]$ rad/s) is used in RFSA with 0.5/999 Hz ($\pi/999$ rad/s) as a step of a frequency grid ($J=1000$ frequencies) and the maximum preset order of a model is $M_{max}=4$. The results shown in Fig. 2 are obtained from uniformly sampled data ($T=1$s). The SNR range [–20, 40] dB is considered. EVT penalization (7) is used with confidence level $\alpha =0.5\%$. The RFSA correctly selects the model order ($M=1$) in the entire range of SNR and attained the CRB at threshold SNR≈–12dB. Maximum execution time of RFSA for a single trial was 12.8 s. The MSE reported by So et al. (2005) is given in the shorter SNR range [–10, 40] dB and attained the CRB over the entire range for the WLS initiated by FFT. The results obtained by WLS method when initiated by RPHD are considerably worse.

Finally the estimation of the frequencies in the three tone case $y=2^{0.5}\sin(0.3\pi)+2^{-0.5}\sin(0.34\pi)+2^{-0.5}\sin(0.7\pi)$ is considered. The SNR values are varied in the range [–10, 40] dB. For each SNR value 1000 MC trials are performed. The frequency interval $f\epsilon[0,0.5]$ Hz ($f\epsilon[0,\pi]$ rad/s) is used in RFSA with 0.5/19 Hz ($\pi/19$ rad/s) as a step of a frequency grid ($J=20$ frequencies). In this particular case the RFSA tends to overestimate the model order. To prevent overestimation we set the maximum order of a model equal to the actual number of sinusoids in a model ($M_{max}=3$) what can be considered equivalent to the condition when the number of sinusoids is known *a priori*. The MSE for the lowest frequency (0.3 rad/s), obtained from uniformly sampled data ($T=1$s) with $K=20$ samples, is shown in Fig 3. EVT penalization (7) is used with confidence level $\alpha =0.5\%$. From Fig 3 it can be seen that RFSA attains CRB at significantly lower SNR threshold than the WLS methods. Almost identical results have been obtained for other two frequencies ($0.34\pi$ and $0.7\pi$). Maximum execution time of a single trial was 0.098 s. From Figs 1–3 it can be concluded that under the same conditions RFSA achieved frequency estimation accuracy at least equal to or better than the accuracy reported by So et al. (2005).

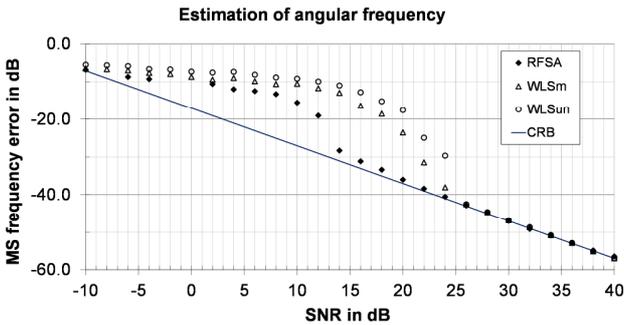

Fig. 3. MSE of the frequency versus SNR obtained by RFSA from irregular data sequences ($K=20$) using EVT model order estimator (7) with $\alpha=0.5\%$ and 20 trial frequencies.

### B. Two sinusoids: one represented by a fraction of its cycle the other one under-sampled

To illustrate the other possibilities of time series analysis by RFSA let us consider an irregularly sampled 1-d signal consisting of $M=2$ superimposed sinusoidal components where the sampled data (64 samples) represent a fraction of the cycle of the first sinusoid whereas the second sinusoid can be considered under-sampled as the time interval between any two adjacent samples is always longer than the full period of the sinusoid. The frequencies of the sinusoids are $f_1=0.011$ Hz and $f_2=2.2$ Hz and their amplitudes are: $A_1=1$ and $A_2=0.5$. The sampling times are calculated by $t_{k+1}=t_k+\tau_{k+1,k}$; $k=1,2,…,64$, where $t_1=1$s and the sampling intervals are uniformly and independently distributed over the interval $\tau_{k+1,k} \epsilon[0.5, 1.5]$ s with mean $\bar{\tau}_{k+1,k} \approx 1$s. Note that minimum sampling interval equals 0.5 s and the Nyquist frequency ≤1 Hz is expected. The phases are $\varphi_1=0$ rad and $\varphi_2=\pi/3$ rad and the additive noise is white and normally distributed with zero mean. *SNR* values in the range [–5, 22] dB are considered in the experiment. The frequency interval $f\epsilon[0,4]$ Hz ($f\epsilon[0,8\pi]$ rad/s) is used in RFSA with 4/511 Hz ($2\pi/511$ rad/s) as a step of the frequency grid (512 frequencies). A refined frequency grid enables precise estimation of very low frequencies. Maximum preset order of a model is $M_{max}=5$. For each SNR value 500 Monte Carlo (MC) trials are performed. Note that the corresponding sampling pattern and additive noise are randomized in each new MC trial. The RFSA sequentially estimates the RFS parameters of most dominant sinusoids and uses MAP and EVT estimators with the corresponding penalization terms (6) and (7) to estimate the number of sinusoids in a model. Given the sampling times and the estimated RFS parameters, the corresponding amplitudes and phases of the sinusoids can be calculated by following the procedure outlined in Appendix C. Each sinusoid can be easily reconstructed by (9) using its RFS parameters or by calculating its frequency, amplitude and phase (Appendix C).

Fig 4 shows an estimated probability of correct model order selection ($M=2$) by RFSA for MAP and EVT estimators with respect to SNR. Fig 5 shows a MSE of two angular frequencies estimated by RFSA from 500 Monte Carlo trials by using MAP model selection criteria (6). Almost identical chart is obtained by using EVT (7) with $\alpha=0.5\%$. From Fig 5 it can be concluded that the threshold for the frequency 0.011 Hz (MSE1) is at SNR≈2 dB and for the frequency 2.2 Hz (MSE2) at SNR≈9 dB. Above these thresholds the MSE approaches to CRB. Note that SNR values in Fig 5 correspond to the 1st sinusoid ($A_1=1$) and are actually lower for about 6 dB for the 2nd sinusoid ($A_2=0.5$). The RFSA estimates both the model order and the frequencies. From Fig 4 it can be seen that correct model selection begins above SNR≈8 dB for model selection criteria MAP (6) and EVT (7) with confidence level $\alpha$ =0.5%. Maximum execution time of a single trial was 1.46 s.

Fig 6 illustrates an instance of irregular sampling pattern with the corresponding data embedded in noise (SNR= 7 dB). Time frame begins at 1 s and ends with 65.10 s. The RFS parameters of the correct model, obtained by the RFSA from data sequence (Fig 6) using MAP estimator (6), are given in Table II. Table III shows true amplitudes, frequencies and phases (subscript $T$), and the corresponding estimates (subscript $E$), obtained from RFS parameters (Table II) in accordance with Appendix C. Evidently, the RFSA is able to precisely estimate the low frequency (0.011 Hz) represented by a fraction of its cycle and the under-sampled frequency (2.2 Hz), which is higher than the Nyquist frequency (1 Hz) based on the minimum sampling interval (0.5 s).

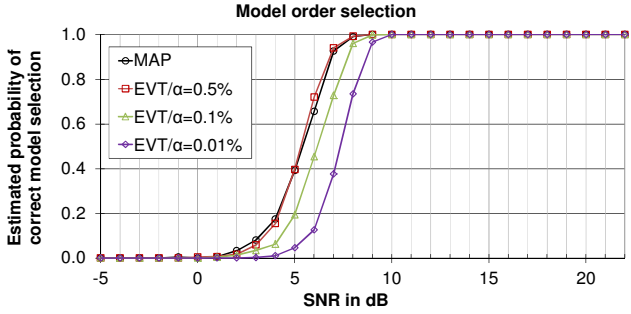

Fig. 4. Estimated probability of correct model order selection ($M$=2) obtained by MC simulations (500 MC trials per SNR value) of irregular data sequence ($K$=64) representing two sinusoids: one with incomplete cycle and the other one under-sampled.

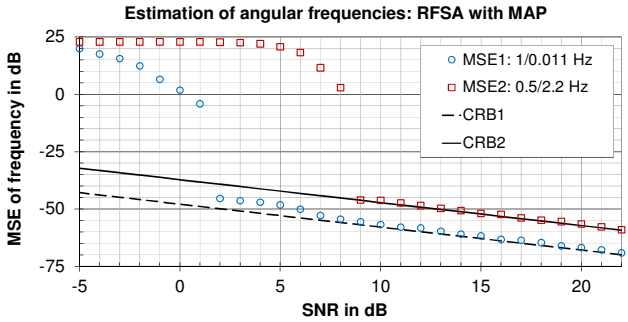

Fig. 5. MSE of the frequencies of sinusoids versus SNR obtained by RFSA from irregular data sequences ($K$=64) using MAP model order estimator (6). Circles and squares represent the MSEs and solid and dashed lines represent the CRBs for the corresponding frequencies.

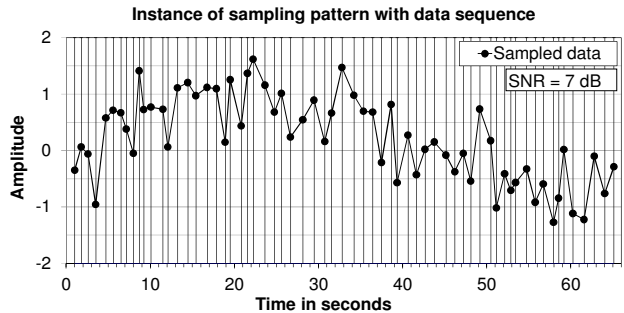

Fig. 6. An instance of irregular sampling pattern ($K$=64) illustrated by vertical lines with data sequence (dots) representing two sinusoids embedded in noise, SNR=7 dB.

TABLE II
ESTIMATED RFS PARAMETERS OF THE MODEL

| Component | 1 | 2 |
|---|---|---|
| $\omega$ [rad] | 6.8901E–02 | 1.3822E+01 |
| $y_1$ | 4.7207E–02 | 3.1457E–01 |
| $y_2$ | 9.5103E–02 | 3.0437E–01 |

TABLE III
TRUE AND ESTIMATED PARAMETERS OF SINUSOIDS

| Component | 1 | 2 |
|---|---|---|
| $A_T$ | 1 | 0.5 |
| $f_T$ [Hz] | 0.011 | 2.2 |
| $\varphi_T$ [rad] | 0 | $\pi/3$ |
| $A_E$ | 9.0025E–01 | 5.1729E–01 |
| $f_E$ [Hz] | 1.0966E–02 | 2.1998E+00 |
| $\varphi_E$ [rad] | –1.6439E–02 | 1.2324E+00 |

*C. Two closely spaced sinusoids with linear trend*

A data sequence consists of a trend line $\kappa t_k + o$ with fixed y–intercept $o$=0.5 and a slope $\kappa$=0.006 $s^{-1}$ and 2 sinusoidal components $A_m \sin(2\pi f_m t + \varphi_m)$, $m$=1,2 with the corresponding amplitudes $A_1 = A_2 = 1$, phases $\varphi_1$=0 rad and $\varphi_2 = \pi/4$ rad and frequencies $f_1$=0.2 Hz and $f_2$=0.2+1/$K$ Hz, where $K$ is the number of samples in the sequence. The additive noise is white and normally distributed with zero mean. The data sequences consisting of $K=64$ and $K=128$ sampling points are analyzed with the corresponding sampling patterns fixed in all MC trials. The SNR interval $SNR \in [-5, 22]$ dB is used for data sequence consisting of K=64 sampling points and $SNR \in [-8, 12]$ dB for K=128. For each SNR value 500 MC trials are performed. In each MC trial a new instance of additive noise is generated. By following the Poisson process, the sampling intervals are exponentially distributed (parameter $\lambda = 0.1$ s$^{-1}$) with mean $1/\lambda = 10$ s. The sampling times are round off to 10 decimals. Maximum preset order of a model is $M_{max}=6$.

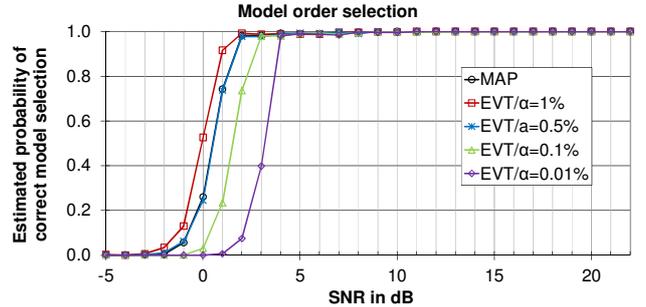

Fig. 7. Estimated probability of correct model order selection ($M$=3) obtained by MC simulations (500 MC trials per SNR value) of irregular data sequence ($K$=64) consisting of two closely spaced sinusoids with linear trend.

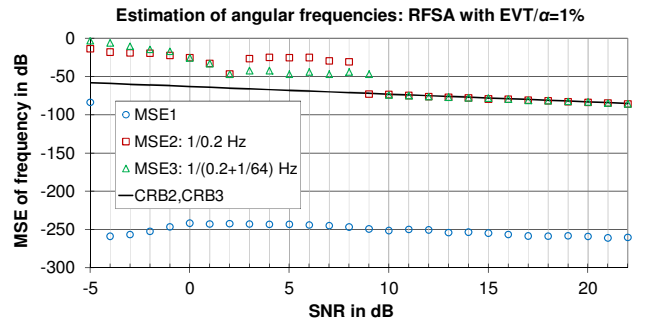

Fig. 8. MSE of the frequencies of sinusoids versus SNR obtained from irregular data sequences ($K$=64) by RFSA using EVT model order estimator (7) with $\alpha$=1%. Circles, squares and triangles, represent the corresponding MSEs and solid line represents the CRB for two closely spaced frequencies.

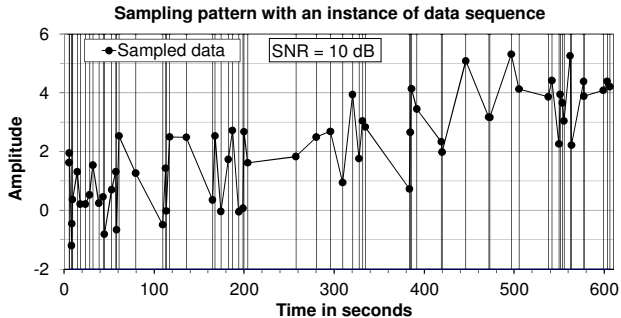

Fig. 9. A sampling pattern ($K=64$) with an instance of a data sequence with zero mean normally distributed additive noise (SNR=10 dB). Vertical lines illustrate the sampling pattern with exponentially distributed intervals.

TABLE IV
ESTIMATED RFS PARAMETERS OF A MODEL

| Component | 1 (Slope) | 2 (0.2Hz) | 3 (0.2+1/64Hz) |
|---|---|---|---|
| $\omega$ [rad] | 2.7923E–13 | 1.2566E+00 | 1.3550E+00 |
| $y_1$ | 5.1969E–01 | 5.1281E–01 | 9.2444E–01 |
| $y_2$ | 5.2126E–01 | 7.5561E–01 | 7.8079E–01 |

TABLE V
CALCULATED PARAMETERS OF SINUSOIDAL COMPONENTS

| Component | 1 (Slope) | 2 (0.2Hz) | 3 (0.2+1/64Hz) |
|---|---|---|---|
| $f$ [Hz] | 4.4441E–14 | 2.0000E–01 | 2.1565E–01 |
| $A$ | 2.1649E+10 | 9.8512E–01 | 9.5813E–01 |
| $\varphi$ [rad] | 2.2495E–11 | 3.3496E–02 | 7.9115E–01 |

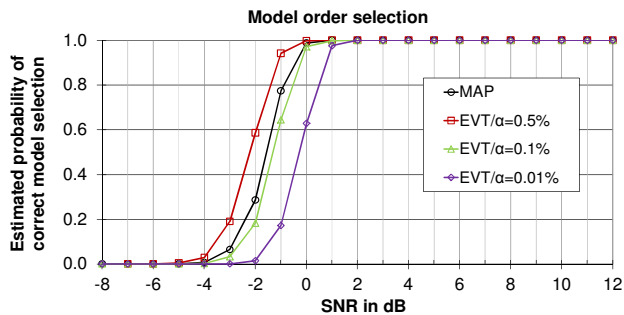

Fig. 10. Estimated probability of correct model order selection ($M=3$) obtained by MC simulations (500 MC trials per SNR value) of irregular data sequence ($K=128$) representing two closely spaced sinusoids with slope.

The frequency interval $f \in [0, 0.5]$ Hz ($\omega \in [0, \pi]$ rad/s) is used with 1/255 Hz ($2\pi/255$ rad/s) as a step of the frequency grid in RFSA (256 frequencies). Note that the correct model order in this experiment is 3 because the RFSA is representing a straight line by a segment of a sinusoid having a frequency of oscillation very close or equal to zero. Depending on the values of the estimated frequencies, (9) or (12) can be used to reconstruct any component from its RFS parameters (radian frequency and two initial samples) returned by RFSA.

Fig 7 shows an estimated probability of correct model order selection ($M$=3) for a data sequence consisting of 64 sampling points obtained from RFSA by using MAP (6) and EVT (7) estimators. For each SNR, 500 MC trials are performed. From Fig 7 it can be seen that very high probability of correct model order selection ($\geq 0.986$) begins at SNR=2 dB but correct model order selection with no misses begins at SNR=10 dB. Maximum execution time of a single trial was 1.18 s. Fig 8 shows a MSE of the estimated frequencies obtained by RFSA from 500 MC trials by using EVT (8) model selection criteria with confidence level $\alpha$=1%. Solid black line in Fig 8 represents a CRB for the two closely spaced frequencies. MSE1 in Fig 8 denotes the MSE of the trend in data ($\omega \approx 0$) and MSE2 and MSE3 denote the MSE of the frequencies of two closely spaced sinusoids 0.2 Hz and 0.2+1/64 Hz, respectively. From Fig 8 it is evident that a perfect reconstruction of all frequencies occurs at the threshold SNR=10 dB. The RFSA is trying to match the linear trend in data sequence with the corresponding segment of a sinusoid by tuning its RFS parameters. The resulting sinusoid generally has extremely low frequency and huge amplitude and it is not possible to calculate the CRB for that frequency. From Fig 8 it can be seen that the MSE of this extremely low frequency ($\omega \approx 0$ Hz) is more than 170 dB lower than the MSE of two other frequencies and it shows the same CRB trend with respect to SNR.

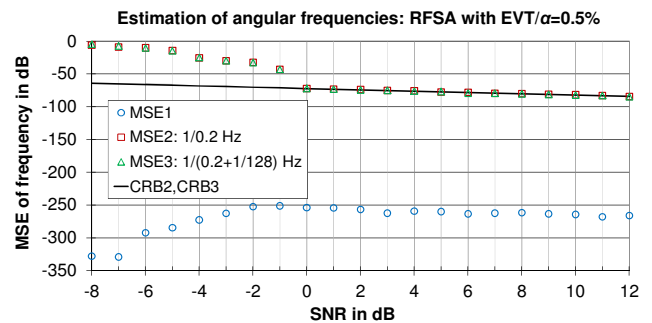

Fig. 11. MSE of the frequencies of sinusoids versus SNR obtained from irregular data sequences ($K=128$) by RFSA using EVT model order estimator (7) with $\alpha$=0.5%. Circles, squares and triangles represent the corresponding MSEs and solid line represents the CRB for two closely spaced frequencies.

Fig 9 illustrates a sampling pattern ($K=64$) and the corresponding data corrupted with noise (SNR=10 dB). The sampling intervals are highly irregular and range from 0.257 s to 53.462 s with mean value 9.533 s. Time frame begins at 5.409 s and ends at 605.994 s. Table IV displays the RFS parameters of a model, estimated from data sequence (Fig 9) by using RFSA with MAP estimator (6). The parameters from Table IV are then used to calculate the corresponding amplitudes and phases of sinusoids, Table V, in accordance with Appendix C. As can be seen from Table V, the linear trend in data is approximated by a segment of a sinusoid having extremely low frequency (4.4441E–14 Hz) and huge amplitude (2.1649E+10). The frequencies of the two closely spaced sinusoids are estimated with negligible errors.

Fig 10 shows an estimated probability of correct model order selection ($M$=3) obtained from MC simulations of data sequence consisting of $K$=128 sampling points by using RFSA with MAP and EVT estimators. For each SNR value in the range [–8,12] dB a 500 Monte Carlo trials are performed. Maximum execution time of a single trial was 2.12 s. Fig 11 shows a MSE of the estimated frequencies obtained by RFSA from 500 MC trials by using EVT (7) model selection criteria with confidence level $\alpha$=0.5%. Again, MSE1 in Fig 11 denotes the MSE of trend in data ($\omega$=0). From Fig 11 it can be seen that a perfect reconstruction of all frequencies occurs above the threshold SNR=0 dB, which is about 10 dB lower than in the previous case ($K$=64).

### D. Three closely spaced sinusoids

An irregular data sequence ($K$=128 and $K$=512 samples) consists of $M$=3 sinusoidal components with frequencies 0.2, 0.2+1/$K$ and 0.2+2/$K$ Hz, amplitudes 1, 0.56234 and 1, and phases 0, $\pi$/4 and $\pi$/3 rad, respectively. Note that the middle sinusoid is 5 dB weaker than the other two. The sampling

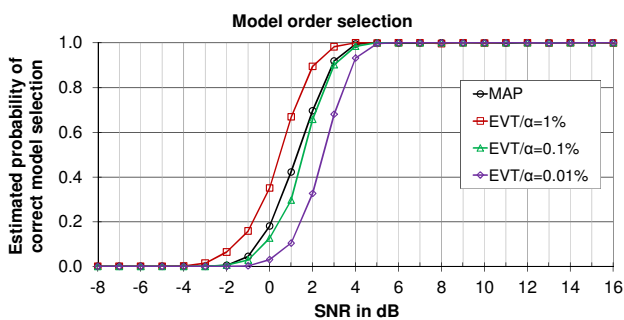

Fig. 12. Estimated probability of correct model order selection ($M$=3) obtained by MC simulations (500 MC trials per SNR value) of irregular data sequences ($K$=128) representing three closely spaced sinusoids.

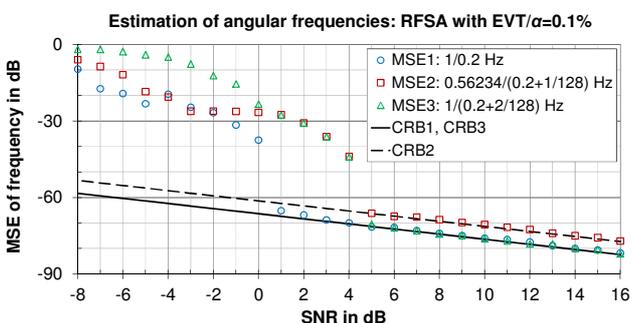

Fig. 13. MSE of the frequencies of sinusoids versus SNR obtained from irregular data sequences ($K$=128) by RFSA using EVT model order estimator (7) with $\alpha$=0.1%. Circles, squares and triangles represent the MSEs and solid and dashed lines represent the CRBs for the corresponding frequencies.

times are calculated by $t_{k+1}=t_k+\tau_{k+1,k}$, where $t_1$=1s and the sampling intervals are uniformly and independently distributed over the interval $\tau_{k+1,k} \in$[0.01,9.99] s with mean $\bar{\tau}_{k+1,k} \approx 5$ s. Note an extremely wide dynamic range of sampling intervals. The additive noise is white and normally distributed with zero mean. The SNR interval $SNR \in [-8,16]$ dB for $K$=128 and $SNR \in [-12,7]$ dB for $K$=512 is used with

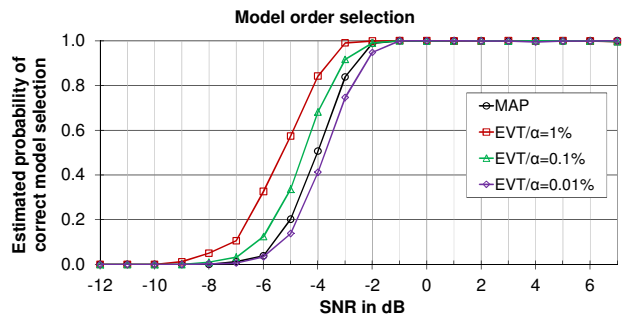

Fig. 14. Estimated probability of correct model order selection ($M$=3) obtained by MC simulations (500 MC trials per SNR value) of irregular data sequences ($K$=512) representing three closely spaced sinusoids.

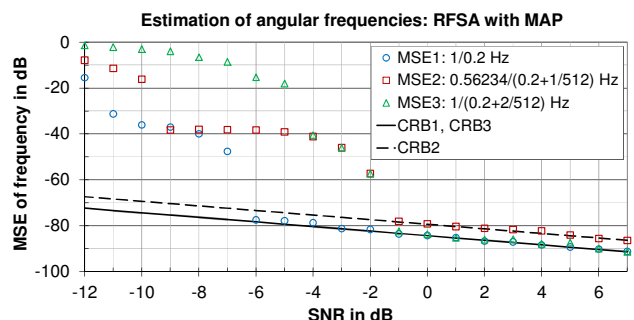

Fig. 15. MSE of the frequencies of sinusoids versus SNR obtained from irregular data sequences ($K$=512) by RFSA using MAP model order estimator (6). Circles, squares and triangles represent the MSEs and solid and dashed lines represent the CRBs for the corresponding frequencies.

1 dB as a step of the noise grid. For each SNR value 500 MC trials are performed. In each MC trial a new sampling pattern and additive noise are generated. The frequency interval $f \in [0,0.5]$ Hz ($\omega \in [0,\pi]$ rad/s) is used with 1/255 Hz (2$\pi$/255 rad/s) as a step of the frequency grid in case of $K$=128 and 1/1023 Hz (2$\pi$/1023 rad/s) in case of $K$=512. Maximum preset order of a model is $M_{max}$=6.

Fig 12 shows an estimated probability of correct model order selection ($M$=3) for a data sequence consisting of 128 sampling points obtained from RFSA by using MAP and EVT estimators and 500 MC trials per each SNR. Maximum execution time of a single trial was 2.57 s. Fig 13 shows a MSE of the frequencies estimated by RFSA from 500 MC trials using EVT (7) model selection criteria with confidence level $\alpha$=0.1%. Fig 14 shows an estimated probability of correct model order selection ($M$=3) for a data sequence consisting of 512 sampling points obtained from RFSA by using MAP and EVT estimators. Maximum execution time of a single trial was 25.94 s. Fig 15 shows a MSE of the frequencies estimated by RFSA from 500 MC trials using MAP (6) model selection criteria. From Figs 12–15 it can be seen that the proposed algorithm, based on the combination of

RFS and parametric methods for model order selection, enables to detect the correct number of sinusoids and to perfectly retrieve the corresponding frequencies from signals highly contaminated with noise.

*E. Time series consisting of ten sinusoids*

An irregular data sequence ($K=200$ samples) consists of $M=10$ sinusoidal components $\{A_m\sin(\omega_m t+\varphi_m), m=1,\ldots,10\}$. Table VI displays the corresponding amplitudes, frequencies and phases uniformly and independently distributed over the intervals $A_m\epsilon[1,2]$, $\omega_m\epsilon[0,1]$ rad/s and $\varphi_m\epsilon[-\pi,\pi]$ rad, respectively. The sampling times are calculated by $t_{k+1}=t_k+\tau_{k+1,k}$; where $t_1=0$ s and the sampling intervals are uniformly and independently distributed over the interval $\tau_{k+1,k}\epsilon[0.01,1.99]$ s with mean $\bar{\tau}_{k+1,k}\approx 1$ s. The additive noise is white and normally distributed with zero mean. The SNR interval SNR$\epsilon[-8,12]$ dB is used with 1 dB as a step of the noise grid. For each SNR value 500 MC trials are performed. In each MC trial a completely new instances of sampling pattern and additive noise are randomly generated. Maximum preset order of a model is $M_{\max}=13$. The frequency interval $f\in[0,0.5]$ Hz ($\omega\in[0,\pi]$ rad/s) is selected with 200 trial frequencies equidistantly distributed over the interval with a step of a frequency grid equal to 1/199 Hz ($2\pi/199$ rad/s).

TABLE VI
TRUE PARAMETERS OF SINUSOIDAL COMPONENTS

| $m$ | $\omega_m$ [rad/s] | $A_m$ | $\varphi_m$ [rad] |
|---|---|---|---|
| 1 | 0.086 | 1.133 | 1.556 |
| 2 | 0.147 | 1.994 | 0.974 |
| 3 | 0.253 | 1.155 | –2.841 |
| 4 | 0.324 | 1.270 | 0.593 |
| 5 | 0.509 | 1.896 | 2.252 |
| 6 | 0.571 | 1.479 | –2.012 |
| 7 | 0.632 | 1.940 | 1.535 |
| 8 | 0.714 | 1.643 | –0.187 |
| 9 | 0.831 | 1.009 | –0.957 |
| 10 | 0.992 | 1.246 | –2.513 |

Fig 16 shows an estimated probability of correct model order selection ($M=10$) obtained from RFSA by using MAP (6) and EVT (7) estimators. Maximum execution time of a single trial was 21.20 s. Fig 17 shows a MSE of the frequencies estimated by RFSA from 500 MC trials using EVT (7) model selection criteria with confidence level $\alpha=0.5\%$. Fig 18 illustrates a MSE of a model, obtained from RFSA by using MAP (6) and EVT (7) estimators, with respect to clean signal (Table VI). Figs 16–18 demonstrate how RFSA can select the correct model order and precisely estimate the corresponding parameters of the significant number of sinusoids heavily contaminated with noise. Recall that SNR (36) in Fig 16–18 corresponds to $A=1$ and, depending on the amplitudes given in Table IV, the SNR is actually higher for at least 0.08 dB ($A_9=1.009$) to maximum 6 dB ($A_2=1.994$).

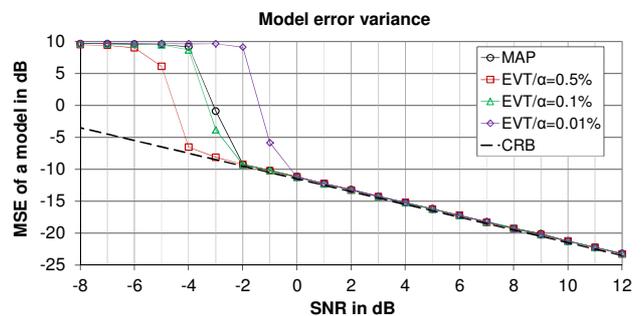

Fig. 18. Estimated MSE of a model obtained by MC simulations (500 MC trials per SNR value) of irregular data sequence ($K=200$) representing 10 sinusoids embedded in noise. Dashed line represents the corresponding CRB.

*F. Some general remarks*

The proposed method is computationally intensive and is best suited for off-line analysis of high frequency and sparse sinusoidal signals. In case of low frequency signals (astronomical observations, electrical biosignals) it can perform analysis in real-time. A comprehensive simulations show that RFSA works fine with any EDC-type model selection criteria but generally achieves the most consistent results using EVT-based model-complexity penalization (7) with confidence level $\alpha=0.5\%$. More details on selecting a proper confidence level can be found in (Nadler & Kontorovich, 2011). It is worth noting that RFSA needs no LM optimization if the frequencies of sinusoidal components coincide with the corresponding trial frequencies, but this condition is uncommon to majority of applications. In order to decrease the probability of LM optimization to get stuck in false local minima it is generally advisable to decrease the step of the frequency grid and in this way bring some trial frequencies close enough to true frequencies.

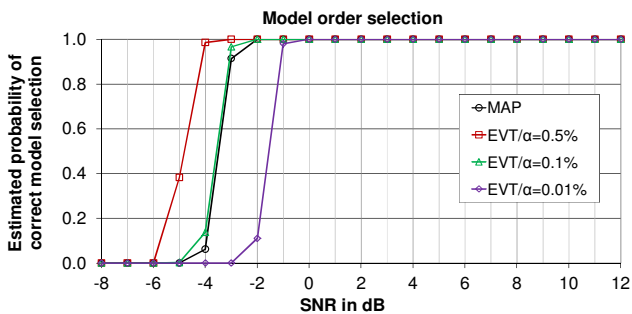

Fig. 16. Estimated probability of correct model order selection ($M=10$) obtained by MC simulations (500 MC trials per SNR value) of irregular data sequence ($K=200$) representing 10 sinusoids.

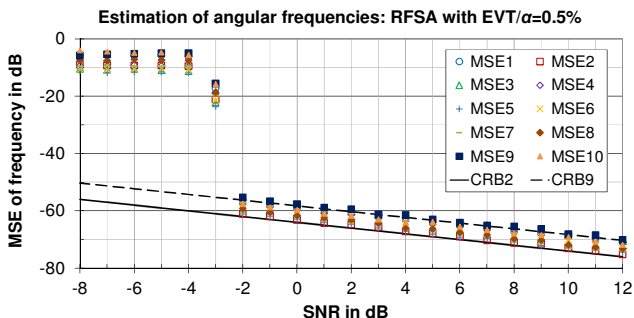

Fig. 17. Estimated MSE of the frequencies of sinusoids versus SNR obtained from irregular data sequences ($K=200$) by RFSA using EVT model order estimator (7) with $\alpha=0.5\%$. Colored markers represent the MSEs of the corresponding frequencies and black lines, solid and dashed, represent only the lowermost ($\omega_2$) and the uppermost ($\omega_9$) CRB, respectively.

IV. CONCLUSION

The recursive formulation of sinusoid can be efficiently combined with common criteria for model order selection in analysis and modeling of nonuniform data sequence representing a sinusoidal signal in noise. By optimizing the parameters of near-optimal sinusoids estimated from the predefined set of trial frequencies, the proposed approach enables decomposition of 1-d signal into a sparse set of sinusoids. It can efficiently model a time series with linear trend in data since a recursive formulation of a straight line can be considered a special case when the frequency of a sine wave approaches zero. It was demonstrated how the proposed algorithm enables to retrieve the under-sampled sinusoid and the sinusoid represented by a fraction of its cycle.

The comprehensive simulations of decomposition of artificial sinusoidal signals corrupted with additive white noise with zero mean always ended by the correct model order selection and by the least-squares estimates of frequencies achieving the Cramer–Rao bound above the threshold signal-to-noise ratio. A relatively high computational complexity can be significantly reduced by parallelizing the execution of (20) for all trial frequencies.

In case of equidistant sampling the computational complexity of the method is further reduced by using Chebyshev's multiple angle formula. A preliminary research also shows that slightly modified recursive formulation of a sinusoid enables to retrieve exponentially damped sinusoids as well.

APPENDIX A

RECURSIVE FORMULATION OF A SINUSOID

Let as consider a three irregularly spaced samples of a sine wave:

$$y_i = A\sin(\omega t_i + \varphi),  \tag{A1}$$

$$y_j = A\sin(\omega t_j + \varphi) = A\sin(\omega t_i + \varphi + \omega \tau_{j,i}) \tag{A2}$$

and

$$y_k = A\sin(\omega t_k + \varphi) = A\sin(\omega t_i + \varphi + \omega \tau_{k,i}), \tag{A3}$$

where $A$ denotes the corresponding amplitude, $\omega$ is the radian frequency $\varphi$ is the phase in radians, $t_k$ is a time point (timestamp) of the corresponding $k$th sample $y_k$, and $\tau_{k,i} = t_k - t_i$ is the difference between the timestamps of $k$th and $i$th sample.

After rewriting (A2) by using basic trigonometric equations we obtain

$$\begin{aligned} y_j &= A\sin(\omega t_i + \varphi)\cos(\omega \tau_{j,i}) + \\ &\quad + A\cos(\omega t_i + \varphi)\sin(\omega \tau_{j,i}) \\ &= \cos(\omega \tau_{j,i})y_i + \\ &\quad + A\cos(\omega t_i + \varphi)\sin(\omega \tau_{j,i}) \end{aligned} \tag{A4}$$

and finally

$$A\cos(\omega t_i + \varphi) = \frac{y_j - \cos(\omega \tau_{j,i})y_i}{\sin(\omega \tau_{j,i})}, \tag{A5}$$

After rewriting (A3) we obtain

$$\begin{aligned} y_k &= A\sin(\omega t_i + \varphi)\cos(\omega \tau_{k,i}) + A\cos(\omega t_i + \varphi)\sin(\omega \tau_{k,i}) \\ &= \cos(\omega \tau_{k,i})y_i + A\cos(\omega t_i + \varphi)\sin(\omega \tau_{k,i}) \end{aligned}, \tag{A6}$$

Substituting (A5) for $A\cos(\omega t_i+\varphi)$ in (A6) and after arrangement we obtain a predictive recurrence relation of a sine wave:

$$y_k = \frac{\sin(\omega \tau_{k,i})}{\sin(\omega \tau_{j,i})} y_j - \frac{\sin(\omega \tau_{k,j})}{\sin(\omega \tau_{j,i})} y_i. \tag{A7}$$

Note that (A7) is independent on how the samples are encountered providing that the time differences, $\tau_{k,j}$, $\tau_{k,i}$, and $\tau_{j,i}$ are used with the corresponding signs. Any sine wave sample can be predicted from any two known samples by knowing their points in time from which the corresponding angular positions in radians can be calculated for the given radian frequency $\omega$. If we set $y_i = y_1$ and $y_j = y_2$ in (A7), an arbitrary sample in a sequence of sine wave samples can be related to the first two samples by:

$$y_k = \frac{\sin(\omega \tau_{k,1})}{\sin(\omega \tau_{2,1})} y_2 - \frac{\sin(\omega \tau_{k,2})}{\sin(\omega \tau_{2,1})} y_1. \tag{A8}$$

APPENDIX B

SIMULTANEOUS LINEAR EQUATIONS FOR THE CALCULATION OF INITIAL SAMPLES OF SINE WAVES

To solve error minimization problem (18), for initial samples of the sine waves $\mathbf{a}$, $\{\mathbf{a}_m\}_{m=1}^{M} = \{y_{m,1}, y_{m,2}\}_{m=1}^{M}$, with known frequencies, $\{\omega_m\}_{m=1}^{M}$, or $x$ parameters, $\{x_m\}_{m=1}^{M}$, the following Jacobian matrix $\mathbf{J}$, parameter vector $\mathbf{a}$ and time series vector $\mathbf{w}$:

$$\mathbf{J} = \begin{bmatrix} b_{1,1} & a_{1,1} & \cdots & b_{M,1} & a_{M,1} \\ b_{1,2} & a_{1,2} & \cdots & b_{M,2} & a_{M,2} \\ \vdots & \vdots & \ddots & \vdots & \vdots \\ b_{1,K-1} & a_{1,K-1} & \cdots & b_{M,K-1} & a_{M,K-1} \\ b_{1,K} & a_{1,K} & \cdots & b_{M,K} & a_{M,K} \end{bmatrix}, \quad \boldsymbol{\alpha} = \begin{bmatrix} y_{1,1} \\ y_{1,2} \\ \vdots \\ y_{M,1} \\ y_{M,2} \end{bmatrix} \text{ and}$$

$$\mathbf{w} = \begin{bmatrix} w_1 \\ w_2 \\ \vdots \\ w_{K-1} \\ w_K \end{bmatrix}.$$

are substituted in (20) to obtain a set of simultaneous linear equations

$$\begin{bmatrix} \sum_{k=1}^{K} b_{1,k}^2 & \cdots & \sum_{k=1}^{K} a_{M,k} b_{1,k} \\ \sum_{k=1}^{K} b_{1,k} a_{1,k} & & \sum_{k=1}^{K} a_{M,k} a_{1,k} \\ \vdots & \ddots & \vdots \\ \sum_{k=1}^{K} b_{1,k} b_{M,k} & \cdots & \sum_{k=1}^{K} a_{M,k} b_{M,k} \\ \sum_{k=1}^{K} b_{1,k} a_{M,k} & \cdots & \sum_{k=1}^{K} a_{M,k}^2 \end{bmatrix} \times \begin{bmatrix} y_{1,1} \\ y_{1,2} \\ \vdots \\ y_{M,1} \\ y_{M,2} \end{bmatrix} = \begin{bmatrix} \sum_{k=1}^{K} b_{1,k} w_k \\ \sum_{k=1}^{K} a_{1,k} w_k \\ \vdots \\ \sum_{k=1}^{K} b_{M,k} w_k \\ \sum_{k=1}^{K} a_{M,k} w_k \end{bmatrix},$$

which can be solved directly for $\boldsymbol{\alpha}$. In case of nonuniform sampling, the coefficients $a_{m,k}$ and $b_{m,k}$ are calculated by (10) and (11) and in case of uniform sampling by (13) and (14).

## APPENDIX C

## CALCULATION OF AMPLITUDE AND PHASE OF A SINUSOID FROM RFS PARAMETERS

Amplitude and phase of a sine wave can be calculated if the corresponding RFS parameters (radian frequency $\omega$ and two samples $y_1$ and $y_2$) are known. A sine wave

$$y = A\sin(\omega t + \varphi) \tag{C1}$$

can be represented by a superposition of the corresponding sine and cosine part. In this way, the two samples of a sine wave obtained at time point $t_1$ and $t_2$ are defined by:

$$y_1 = B\sin(\omega t_1) + C\cos(\omega t_1), \tag{C2}$$

$$y_2 = B\sin(\omega t_2) + C\cos(\omega t_2). \tag{C3}$$

By combining (C2) and (C3) one can calculate the amplitude of the corresponding cosine

$$C = \frac{y_1 \sin(\omega t_2) - y_2 \sin(\omega t_1)}{\sin(\omega t_2)\cos(\omega t_1) - \sin(\omega t_1)\cos(\omega t_2)} \tag{C4}$$

and the sine part

$$B = \frac{y_1}{\sin(\omega t_1)} - C\frac{\cos(\omega t_1)}{\sin(\omega t_1)}. \tag{C5}$$

By using (C4) and (C5) the amplitude of the original sine wave can be calculated by:

$$A = \sqrt{B^2 + C^2}. \tag{C6}$$

The phase of a sine wave can be calculated by using Euler's relation:

$$e^{j\varphi} = C + jB. \tag{C7}$$

In case of uniform sampling the procedure is the same after converting parameter $x$ into the frequency (16).

## APPENDIX D

## A SET OF SIMULTANEOUS LINEAR EQUATIONS TO CALCULATE INCREMENT VECTOR FOR LM OPTIMIZATION OF RFS PARAMETERS

To solve the error minimization problem (21) with respect to increment vector $\boldsymbol{\delta}$, the following parameter vector $\boldsymbol{\beta}$, parameter increment vector $\boldsymbol{\delta}$, time series vector $\mathbf{w}$, time series prediction vector $\mathbf{P}(\boldsymbol{\beta})$ and Jacobian matrix $\mathbf{J}$:

$$\boldsymbol{\beta} = \begin{bmatrix} \omega_1 \\ y_{1,1} \\ y_{1,2} \\ \vdots \\ \omega_M \\ y_{M,1} \\ y_{M,2} \end{bmatrix}, \boldsymbol{\delta} = \begin{bmatrix} \delta\omega_1 \\ \delta y_{1,1} \\ \delta y_{1,2} \\ \vdots \\ \delta\omega_M \\ \delta y_{M,1} \\ \delta y_{M,2} \end{bmatrix}, \mathbf{w} = \begin{bmatrix} w_1 \\ w_2 \\ w_3 \\ \vdots \\ w_K \end{bmatrix}, \mathbf{P}(\boldsymbol{\beta}) = \begin{bmatrix} \sum_{m=1}^{M} y_{m,1} \\ \sum_{m=1}^{M} y_{m,2} \\ \sum_{m=1}^{M} y_{m,3} \\ \vdots \\ \sum_{m=1}^{M} y_{m,K} \end{bmatrix},$$

$$\mathbf{J} = \begin{bmatrix} \frac{\partial y_{1,1}}{\partial \omega_1} & \frac{\partial y_{1,1}}{\partial y_{1,1}} & \frac{\partial y_{1,1}}{\partial y_{1,2}} & \cdots & \frac{\partial y_{M,1}}{\partial y_{M,2}} \\ \frac{\partial y_{1,2}}{\partial \omega_1} & \frac{\partial y_{1,2}}{\partial y_{1,1}} & \frac{\partial y_{1,2}}{\partial y_{1,2}} & \cdots & \frac{\partial y_{M,2}}{\partial y_{M,2}} \\ \frac{\partial y_{1,3}}{\partial \omega_1} & \frac{\partial y_{1,3}}{\partial y_{1,1}} & \frac{\partial y_{1,3}}{\partial y_{1,2}} & \cdots & \frac{\partial y_{M,3}}{\partial y_{M,2}} \\ \vdots & \vdots & \vdots & \ddots & \vdots \\ \frac{\partial y_{1,K}}{\partial \omega_1} & \frac{\partial y_{1,K}}{\partial y_{1,1}} & \frac{\partial y_{1,K}}{\partial y_{1,2}} & \cdots & \frac{\partial y_{M,K}}{\partial y_{M,2}} \end{bmatrix},$$

are substituted in (22) to obtain a set of simultaneous linear equations:

$$\begin{bmatrix} (1+\gamma)\sum_{k=1}^{K}\left(\frac{\partial y_{1,k}}{\partial \omega_1}\right)^2 & \cdots & \sum_{k=1}^{K}\frac{\partial y_{M,k}}{\partial y_{M,2}}\frac{\partial y_{1,k}}{\partial \omega_1} \\ \sum_{k=1}^{K}\frac{\partial y_{1,k}}{\partial \omega_1}\frac{\partial y_{1,k}}{\partial y_{1,1}} & \cdots & \sum_{k=1}^{K}\frac{\partial y_{M,k}}{\partial y_{M,2}}\frac{\partial y_{1,k}}{\partial y_{1,1}} \\ \sum_{k=1}^{K}\frac{\partial y_{1,k}}{\partial \omega_1}\frac{\partial y_{1,k}}{\partial y_{1,2}} & \cdots & \sum_{k=1}^{K}\frac{\partial y_{M,k}}{\partial y_{M,2}}\frac{\partial y_{1,k}}{\partial y_{1,2}} \\ \vdots & \ddots & \vdots \\ \sum_{k=1}^{K}\frac{\partial y_{1,k}}{\partial \omega_1}\frac{\partial y_{M,k}}{\partial y_{M,2}} & \cdots & (1+\gamma)\sum_{k=1}^{K}\left(\frac{\partial y_{M,k}}{\partial y_{M,2}}\right)^2 \end{bmatrix} \times \begin{bmatrix} \delta\omega_1 \\ \delta y_{1,1} \\ \delta y_{1,2} \\ \vdots \\ \delta\omega_M \\ \delta y_{M,1} \\ \delta y_{M,2} \end{bmatrix} =$$

$$= \begin{bmatrix} \sum_{k=1}^{K}\frac{\partial y_{1,k}}{\partial \omega_1}\left(w_k - \sum_{m=1}^{M}y_{m,k}\right) \\ \sum_{k=1}^{K}\frac{\partial y_{1,k}}{\partial y_{1,1}}\left(w_k - \sum_{m=1}^{M}y_{m,k}\right) \\ \sum_{k=1}^{K}\frac{\partial y_{1,k}}{\partial y_{1,2}}\left(w_k - \sum_{m=1}^{M}y_{m,k}\right) \\ \vdots \\ \sum_{k=1}^{K}\frac{\partial y_{M,k}}{\partial \omega_M}\left(w_k - \sum_{m=1}^{M}y_{m,k}\right) \\ \sum_{k=1}^{K}\frac{\partial y_{M,k}}{\partial y_{M,1}}\left(w_k - \sum_{m=1}^{M}y_{m,k}\right) \\ \sum_{k=1}^{K}\frac{\partial y_{M,k}}{\partial y_{M,2}}\left(w_k - \sum_{m=1}^{M}y_{m,k}\right) \end{bmatrix}$$

that can be solved directly for **δ**. Partial derivatives $\frac{\partial y_{m,k}}{\partial \omega_m}$, $\frac{\partial y_{m,k}}{\partial y_{m,1}}$ and $\frac{\partial y_{m,k}}{\partial y_{m,2}}$, can be calculated by (23)–(26). In case of uniform sampling parameter $x$ should be substituted for $\omega$ and the corresponding partial derivatives are calculated by (27)–(29).